\documentclass[pre,twocolumn,amsmath,amssymb,superscriptaddress,showpacs]{revtex4-2}
\usepackage{graphicx}
\usepackage{latexsym}
\usepackage{dcolumn}
\usepackage{bm}
\usepackage{color}
\usepackage{amssymb}
\usepackage[normalem]{ulem}
\usepackage{mathtools}

\newcommand{\be}{\begin{equation}}
\newcommand{\ee}{\end{equation}}
\newcommand{\bea}{\begin{eqnarray}}
\newcommand{\eea}{\end{eqnarray}}
\newcommand{\bse}{\begin{subequations}}
\newcommand{\ese}{\end{subequations}}

\newcommand{\e}{\varepsilon}

\newcommand{\comment}[1]{}

\begin{document}

\title{Relaxation dynamics and finite-size effects in a simple model of condensation}
\date{\today}

\author{Gabriele Gotti}
\affiliation{Dipartimento di Fisica e Astronomia, Universit\`a di Firenze,
via G. Sansone 1 I-50019, Sesto Fiorentino, Italy}
\affiliation{Istituto dei Sistemi Complessi, Consiglio Nazionale
delle Ricerche, via Madonna del Piano 10, I-50019 Sesto Fiorentino, Italy}

\author{Stefano Iubini}
\email{stefano.iubini@cnr.it}
\affiliation{Istituto dei Sistemi Complessi, Consiglio Nazionale
delle Ricerche, via Madonna del Piano 10, I-50019 Sesto Fiorentino, Italy}
\affiliation{Istituto Nazionale di Fisica Nucleare, Sezione di Firenze, 
via G. Sansone 1 I-50019, Sesto Fiorentino, Italy}

\author{Paolo Politi}
\email{paolo.politi@cnr.it}
\affiliation{Istituto dei Sistemi Complessi, Consiglio Nazionale
delle Ricerche, via Madonna del Piano 10, I-50019 Sesto Fiorentino, Italy}
\affiliation{Istituto Nazionale di Fisica Nucleare, Sezione di Firenze, 
via G. Sansone 1 I-50019, Sesto Fiorentino, Italy}

\begin{abstract}
We consider a simple, purely stochastic model characterized by two conserved
quantities (mass density $a$ and energy density $h$) which is known to
display a condensation transition when $h>2a^2$: in the localized phase 
a single site hosts a finite fraction of the whole energy.
Its equilibrium properties in the thermodynamic limit are known and in a recent paper 
(Gabriele Gotti, Stefano Iubini, Paolo Politi, Phys. Rev. E 103, 052133 (2021))
we studied the transition for finite systems. 
Here we analyze finite-size effects on the energy distribution and on the relaxation
dynamics, showing that extremely large systems should be studied in order to observe
the asymptotic distribution and even larger systems should be simulated
in order to observe the expected relaxation dynamics.
\end{abstract}
\maketitle

\section{Introduction}

The phenomenon of homogeneous condensation in the real space is a rather common
mechanism  in many-body statistical systems.   
When it occurs, the usual relaxation towards a uniform (equilibrium) state
does not take place and a macroscopic amount of energy~%
\footnote{In the model studied below the localized physical quantity is the energy.
In other models it might be the mass or a different quantity conserved by the dynamics.} 
remains localized in space for arbitrary 
long times~\cite{Drouffe1998_JPA,Godreche2003_JPA,Majumdar2005_PRL,Evans2005_JPA,Evans2006_JSP,Majumdar2010_LesHouches,Zannetti2014_PRE,Mori2021}.

In the last decade a model of condensation  
has come to the attention of researchers because of its simplicity and its connections with 
physical setups.
Such a model is also related to the Kipnis-Marchioro-Presutti (KMP) model of heat flow~\cite{kmp82}, the first model
where forty years ago it was possible to rigorously prove that heat flows 
according to the diffusion equation. 
KMP is a purely stochastic model defined on a one dimensional lattice of $N$ sites
where a positive, real quantity  $c_i$ is defined on every lattice site $i$,
and whose sum $A=\sum_i c_i \equiv Na$ is conserved by dynamics.
The simplest microcanonical move $(c_i \to c_i',
c_{i+1} \to c_{i+1}')$ involves a pair of neighbouring sites, $(i,i+1)$,
so as to keep constant the sum $c_i + c_{i+1}$.
This can be done in such a way to satisfy detailed balance:
$c_i' = r (c_i + c_{i+1})$ and
$c_{i+1}' = (1-r) (c_i + c_{i+1})$, where $r$ is a random number in the
unitary interval.
The KMP model is simply characterized by diffusion and it certainly cannot
produce any localization: indeed the uniform
spreading of the conserved quantity $A$ is predicted.

Let us now suppose that there exists a second conserved quantity $H$, locally equal to the 
square of the first quantity, $H=\sum_i c_i^2\equiv Nh$. This new model does present
a condensation/localization phenomenon if $h>h_c\equiv 2a^2$, because of the presence of the
additional constraint on $H$~\cite{Szavits2014_PRL}. Accordingly,
in the thermodynamic
limit a \textit{single} site $k$ hosts a finite fraction of the whole energy $H$, namely
$\e_k \equiv c_k^2 = (h-h_c)N$~\cite{Gradenigo2021_JSTAT,Gradenigo2021_EPJE}.
Because of translational invariance any site $k$ is eligible to host the
condensate and according to the details of the dynamic the site $k$ may be either 
fixed or it may diffuse. 
The simplest microcanonical dynamics is a generalization of the KMP one:
a triplet of neighboring sites, $(i-1,i,i+1)$, is randomly chosen and
we perform a uniform move so as to conserve the sum of the masses,
$c_{i-1}+c_i +c_{i+1} = M$, and the sum of the energies,
$c^2_{i-1}+c^2_i +c^2_{i+1} = E$.
This model with two conserved quantites has been called C2C~\cite{GIP21}
(Condensation with 2 Conserved quantities) and above dynamics
has been called MMC (Microcanonical Monte Carlo)~\cite{JSP_DNLS}.

The relevance of the C2C model is also related to its closeness with
the Discrete NonLinear Schr\"odinger (DNLS) equation, commonly employed to describe  wave propagation phenomena in lattices
where nonlinearity is dominant and dissipation is negligible~\cite{kevrekidis09}.
In one dimension, the DNLS dynamics corresponds to the Hamiltonian equation
\be
i\dot{z}_n = -2|z_n|^2 z_n - z_{n+1} - z_{n-1} 
\ee
where the complex amplitudes $z_i$ are defined on each lattice site.
Upon identifying $c_i = |z_i|^2$, conservation of C2C mass corresponds exactly to
the conservation of the DNLS total norm, $\sum_i |z_i|^2 \equiv Na$.
Instead, the DNLS total energy  $\mathcal{H}$ (also conserved) has a more complicated expression
\be\label{eq:dnls_h}
\mathcal{H}=\sum_i |z_i|^4 + (z_i^* z_{i+1} + c.c.)
\ee 
which can not be directly cast in terms of C2C observables due
to the presence of the phase-dependent hopping term $(z_i^* z_{i+1} + c.c.)$.
However, it can be shown~\cite{Gradenigo2021_JSTAT,arezzo22} that close to the critical curve separating the homogeneous and the localized 
phase, the hopping term in Eq.~(\ref{eq:dnls_h}) becomes negligible and 
$\mathcal{H}=\sum_i c_i^2$. For this reason, DNLS and C2C equilibrium critical
behaviors are equivalent.

As already mentioned, the equilibrium state in the condensed phase of the C2C model
with total energy $h N$ displays a single site hosting the energy $(h-h_c)N$ while the remaining
energy $h_c N$ is distributed exponentially in the rest of the system~\cite{Rasmussen2000_PRL}.
However, this statement is true only in the thermodynamic limit $N\to \infty$:
finite size effects are important and become more and more important 
as the critical point is approached, $h\to h_c^+$~\cite{Gradenigo2021_EPJE}.
A systematic study of the influence of finite-size effects on the equilibrium localization properties
of this and other condensation models 
was performed in Ref.~\cite{GIP21}. There it was shown that 
 the order parameter of the transition, 
quantified by means of
 the participation ratio, see Eq.~(\ref{eq.Y2}), 
may display a nonmonotonic dependence on the lattice size $N$.  

In this paper we investigate in detail the role of finite size effects
for the equilibrium energy distribution of the C2C model and we extend our analysis to
its relaxation (i.e. out-of equilibrium) dynamics.
Two different classes of stochastic update rules will be analyzed, namely those corresponding to   local and nonlocal dynamics.
In the former case, the microcanonical MMC move involves a triplet of neighboring
sites $(i-1,i,i+1)$, as discussed above; in the latter case,
the three sites of the triplet, $(i,j,k)$, may be located everywhere.

The paper is organized as follows. In Sec.~\ref{sec.theory} we discuss the expected behaviors for
a system in the thermodynamic limit, both at equilibrium and during the relaxation process. 
In Sec.~\ref{sec.fse} we discuss the role of finite size effects.
We start from the $N-$dependency of the order parameter characterizing the localization 
transition, the so-called participation ratio, which shows a non monotic behavior
with a minimum.
The analysis of the energy distribution (Sec.~\ref{ss.ed}) clarifies that the minimum separates
a non localized regime from the (asymptotic) localized one.
The relaxation to the equilibrium distribution requires much longer times for local dynamics
and this is also visible in the analysis of the relaxation process (Sec.~\ref{ss.rd}):
coarsening triggers later and it is slower, if dynamics is local.
The final Sec.~\ref{sec.conc} contains a discussion of the results and some concluding remarks.

\section{Equilibrium and nonequilibrium properties in the large-$N$ limit}
\label{sec.theory}

In the limit of large system size $N$, the description of the condensation process  
is relatively simple, as it involves a single peak
superposed to a stationary background with known fluctuations.
Assuming such a perfect separation between peaks and background distributions, several important properties of
the condensation dynamics can be  derived analytically, both at equilibrium and for relaxation
problems.

The equilibrium properties of the C2C model have been studied in detail~\cite{Szavits2014_PRL,
Szavits2014_JPA,Gradenigo2021_JSTAT,Gradenigo2021_EPJE,gotti22}.
Here we summarize the main results for the system in the thermodynamic limit and we refer to the
cited bibliography for the details of the calculations.
The model is defined for $h\ge a^2$ and the ground state $h=a^2$ 
corresponds to a perfectly homogeneous configuration,
$c_i =a\;\forall i$.
With increasing $h$ at fixed $a$, spatio-temporal fluctuations increase
until the critical line $h=h_c\equiv 2a^2$ is attained.
Here local masses $c_i$ are distributed exponentially,
as expected by the fact that for $h=2a^2$ the variance of the 
mass equals the square of the average mass.
The region above the critical line corresponds to a condensed
phase because the energy $(h-h_c)N$ is localized on a single size.
This phase can be described in the microcanonical ensemble
but not in the grand canonical one, which is restricted to the
homogeneous phase, $a^2 \le h \le 2a^2$. In such uniform phase
there is ensemble equivalence and the temperature varies
between $T=0$ along the ground state line and $T=\infty$
along the critical line.
The microcanonical temperature above the critical line is
equal to $T=+\infty$ in all the localized phase~\cite{Gradenigo2021_EPJE}.

The order parameter of the condensation transition is the participation ratio 
\be
Y_2(N) \equiv \frac{\langle\, \overline{\e_i^2}\,\rangle}{\displaystyle N\langle\,\overline{\e_i}\,\rangle^2} =
\frac{\langle \,\overline{\e_i^2}\,\rangle}{Nh^2} \,, 
\label{eq.Y2}
\ee
where $\overline{(\cdots )}$ is a spatial average and $\langle\cdots\rangle$ is a statistical average.
The participation ratio allows to discriminate between homogeneous and localized states: in the former case
the second moment of the energy distribution is finite and $Y_2$ vanishes as $1/N$;
in the latter case $\langle\, \overline{\e_i^2}\,\rangle$ diverges with $N$ and $Y_2$
converges to a positive constant which depends on the excess energy $h-h_c$. 
More precisely, if we evaluate $Y_2$ for a single peak sitting on top of an homogeneous background
(i.e. the equilibrium localized state expected for large $N$, see Ref.~\cite{Gradenigo2021_JSTAT}) we obtain
\be
Y_2(N) = \frac{24 a^4}{Nh^2} + \left(\frac{h-h_c}{h}\right)^2 \to \left(\frac{h-h_c}{h}\right)^2 ,
\label{eq.Y2_C2C}
\ee
where we have used the fact that the average value of $\e_i^2= c_i^4$ is equal to $24a^4$ if $c_i$ are
distributed exponentially with average value equal to $a$.

Previous Eq.~(\ref{eq.Y2_C2C}) is a first example of ``finite-size effect" because it provides
an $N-$dependent correction to the asymptotic expression of $Y_2(N)$, but such simple expression is not
valid in general, as shown and discussed in detail in Ref.~\cite{GIP21}.
We will return to this point in the next Section.

\begin{figure}[h!]
\includegraphics[width=0.46\textwidth,clip]{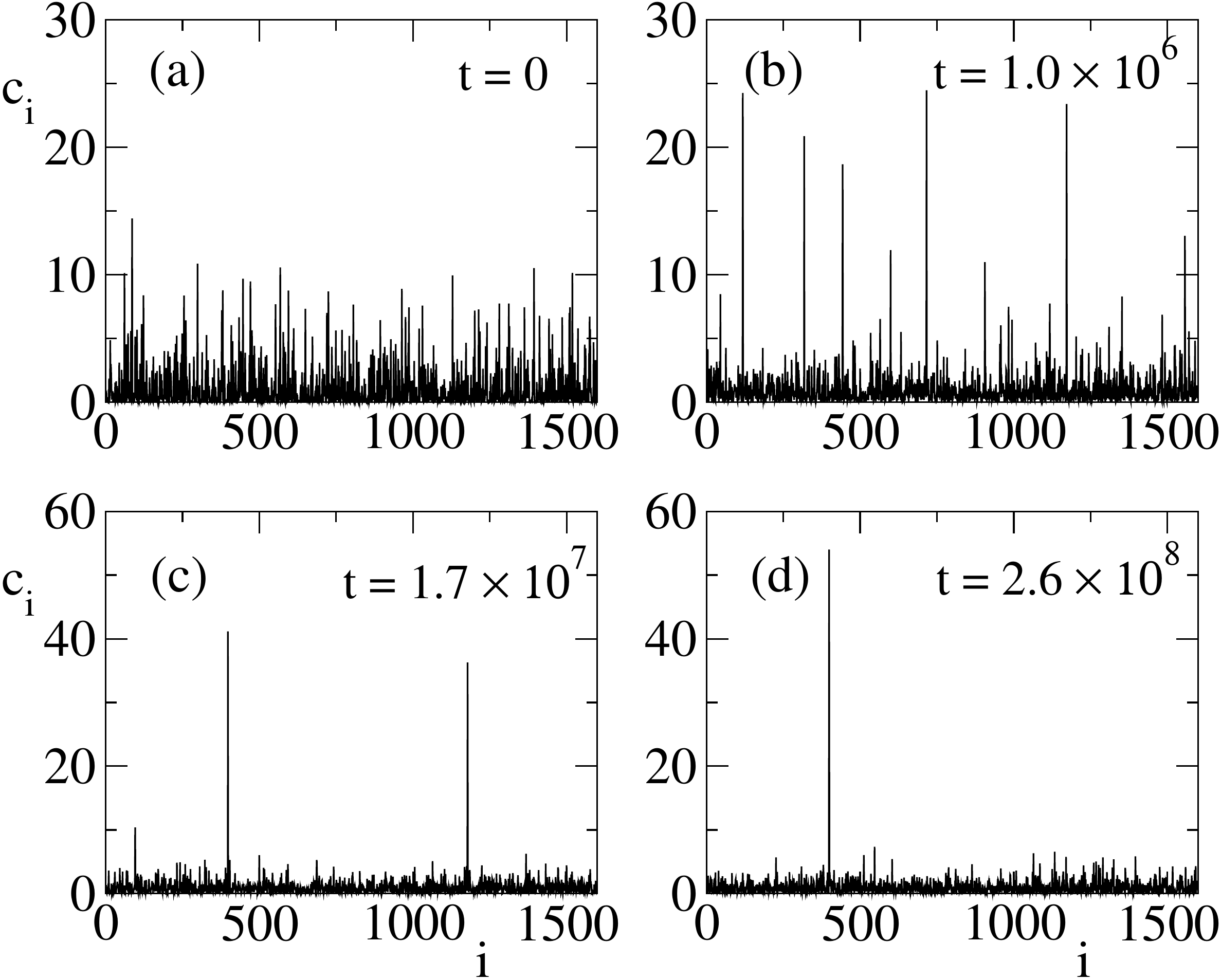}
\caption{The condensation dynamics via coarsening of peaks for a chain with $N=1600$ and periodic
boundary conditions. The initial condition is generated from a superposition  of two exponential
distributions for the $c(i)$'s. The corresponding conserved densities are $a= 1$
and $h= 3.7$ and a local update dynamics is considered.}
\label{fig_3.2}
\end{figure}

The same schematic representation of the condensed phase (single peak plus uniform background) 
can be used to evaluate how the system relaxes to equilibrium.
In Fig.~\ref{fig_3.2} we report some snapshots at different times of the relaxation process
for a system  with local dynamics,  $N=1600$ and $h=3.7$ (here and in the following we set $a=1$ without loss 
of generality~\cite{JSP_DNLS}).   The initial condition belongs to the condensed phase $(h>h_c)$ 
but it is delocalized, see panel (a). Hence, the system is expected to spontaneously localize 
energy as the equilibrium state is approached. 
Indeed, after an initial transient, a generic configuration is made up by a certain number of peaks
(of variable height) on top of a background, see panels (b), (c). Finally, for large enough times,
a single peak is clearly visible in panel (d). 
As explained in Ref.~\cite{JSP_DNLS}, relaxation occurs via a coarsening process where
the number of peaks decreases in time as a power law. This happens because the evolution of a triplet of
masses $c_1,c_2,c_3$ where one mass (e.g., $c_1$) is much larger than the others 
is such that $|c_1' -c_1|$ scales as $1/c_1$, so that the variation of its energy,
$|\e_1'-\e_1|$, is of order one, irrespective of the peak's energy.
Therefore relaxation can be understood as a process where neighbouring peaks
exchange energy packages through the background. 
Because of such \textit{symmetric} exchanges, the energy $\e_p$ of a peak performs
a symmetric random walk. Consequently, one can conclude that the disappearance of a peak requires on average 
a number of order $\e_p^2$ of such exchanges~\cite{JSP_DNLS}.

In order to derive the time dependence of the relaxation process~\footnote{As usual for Monte-Carlo methods, the 
time unit is here defined as the number of Monte-Carlo updates divided by the system size $N$.}, we need to know the typical time $\Delta t$ for
a single exchange to occur and the dependence of the energy $\e_p$ on the peak interdistance $\ell=
1/\rho$, where $\rho$ is the density of peaks.
Since a peak hosts a finite fraction of the energy of $\ell$ sites, $\e_p$ must be proportional to $\ell$.
As for $\Delta t$, we need to distinguish between local and nonlocal dynamics.
In the former case, we can imagine that the energy packet travels diffusively along the background
with the caveat that it may be reabsorbed by the departure peak, therefore nullifying the
potential exchange of energy between neighboring peaks. Since the probability that an emitted package attains the neighboring peak
before being reabsorbed is equal to $1/\ell$, $\Delta t \simeq \ell$ and the total time because
a single peak disappears scales as $\e_p^2 \times\ell \simeq \ell^3$.
This means that that in a multi-peak configuration the average peak interdistance satisfies the relation
$\ell^3(t) \sim t$, i.e. $\ell(t) \simeq t^{1/3}$.
If the dynamics is nonlocal, peaks can interact either directly  or through a
finite number of exchanges, which does not depend on the peak interdistance $\ell$.
For this reason $\Delta t \simeq 1$, the time of disappearance
of a single peak now scales as $\ell^2$, therefore $\ell(t) \simeq t^{1/2}$.

In conclusion, we expect that the relaxation process in the condensed region occurs via a coarsening dynamics, 
$\ell(t) \simeq t^n$, characterized
by exponents $n=1/3$ and $n=1/2$ for local and nonlocal dynamics, respectively.
It is worth stressing that such exponents match the relaxation law of
the one-dimensional Ising model via Kawasaki (spin-exchange) dynamics,
which may be either local or nonlocal~%
\footnote{Dynamics of the Ising model is easier to understand if positive and negative spins are replaced
by particles and holes, because spin-exchange means that a particle hops on a hole, i.e.
an empty site. Relaxation occurs because clusters exchange particles, much in the same way
peaks exchange energy in the C2C model. Local/Nonlocal dynamics means that particles hop on a neighbouring hole
(local dynamics) or in any available hole (nonlocal dynamics).
In this picture it is easy to derive the coarsening exponents for the Ising model:
$n=1/3$ for local dynamics~\cite{livi17} and $n=1/2$ for global one~\cite{RK1996}.}.

\section{Finite size effects}
\label{sec.fse}
When the lattice size $N$ is not large enough to invoke the thermodynamic limit, significant finite-size effects arise.
In this regime, a generic configuration cannot be any more decomposed into two perfectly distinct contributions for peaks (a delta-shaped distribution) and background.
As a result, analytic descriptions become much more difficult~\cite{Mori2021}.
In the following we discuss finite size effects for the equilibrium distribution and
for the relaxation process.
 
\subsection{Energy distributions}
\label{ss.ed}

\begin{figure}[ht!]
\includegraphics[width=0.46\textwidth,clip]{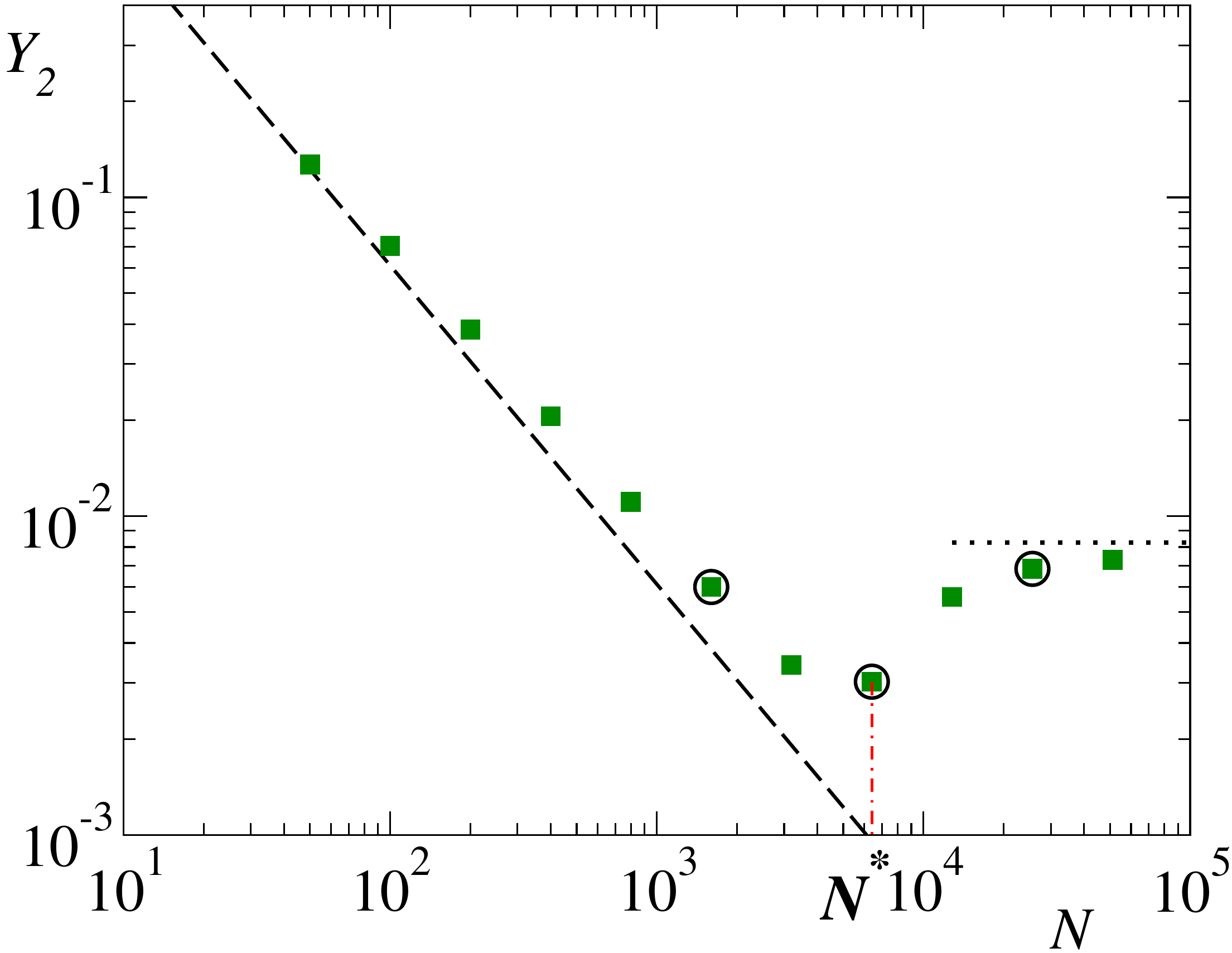}
\caption{The participation ratio for $h=2.2$ and different values of $N$ (filled green squares).
The dashed line is the analytical expression $Y_2=24/Nh^2$ expected to be valid for 
$N \ll N^*$, while the horizontal dotted line is the analytical expression $Y_2 = (h-h_c)^2/h^2$
valid for $N \gg N^*$,  see Eq.~(\ref{eq.Y2_C2C}).
The three symbols enclosed by a circle refer to three values of $N$ for which we study the
energy distributions, see Figs.~\ref{fig.ed} and \ref{fig.ed_lnl}.} 
\label{fig.y2}
\end{figure}

In Fig.~\ref{fig.y2} we plot the participation ratio versus $N$ for a value of $h=2.2 > h_c$, 
close but not too close to the critical value $h_c$. The asymptotic value of $Y_2$ and its $1/N$ correction predicted by
Eq.~(\ref{eq.Y2_C2C}) are shown in dotted and dashed lines, respectively. 
Unexpected finite-size effects manifest themselves in the
appearance of a minimum of $Y_2$ for $N=N^*\simeq 6400$~\footnote{The minimum $Y_2(N^*)$ scales as $(h-h_c)^3$, which means that
$N^* \sim (h-h_c)^{-3}$. See Ref.~\protect\cite{GIP21} for more details.}. 
Such a non-monotonic behavior is forbidden in Eq.~(\ref{eq.Y2_C2C}).

Since the minimum of $Y_2$ separates different behaviors, we have analyzed the stationary energy distribution $P(\e)$
for three different values of the system size, $N=1600$, $N=N^*=6400$, and $N=25600$, 
corresponding to the minimum $N=N^*$, to a smaller size value, and to a larger size value.

\begin{figure}[ht!]
\includegraphics[width=0.46\textwidth,clip]{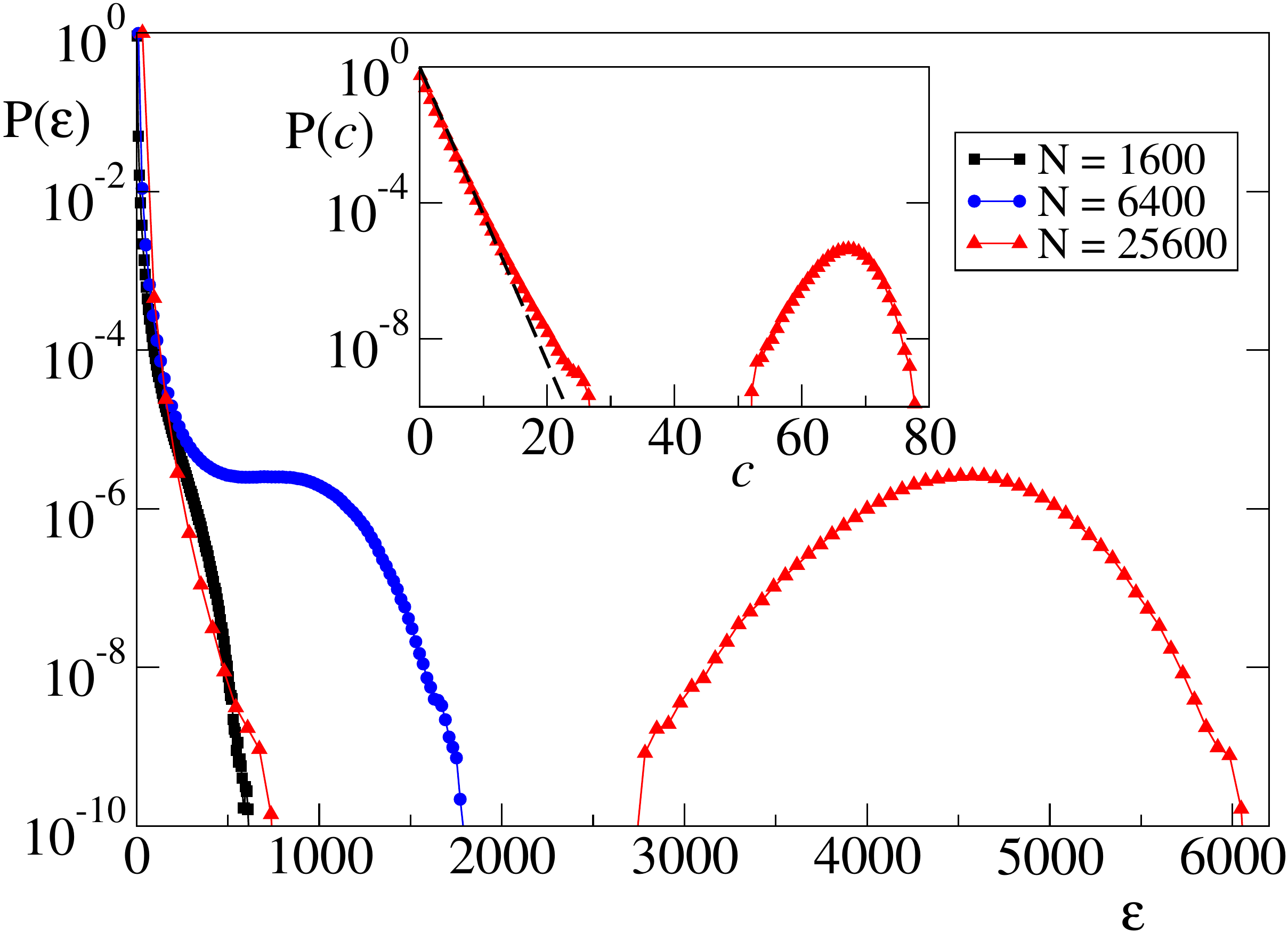}
\caption{Energy distributions for $a=1,h=2.2$, and for the three values of $N$ indicated in the legend box.
Inset: The mass distribution for the largest value of $N$ (same symbols).
The dashed straight line represents the exponential distribution.}
\label{fig.ed}
\end{figure}

Results are shown in Fig.~\ref{fig.ed}. If $N< N^*$, $P(\e)$ is a decreasing function and there is no
indication of a condensate (black squares). When $N=N^*$ a bump appears (blue circles) but only for $N=25600=4N^*$ 
the energy distribution is composed of two separated contributions corresponding to the background  and to the condensate (red triangles). 
Clearly, the condensate distribution for the largest $N$ is far from being a Dirac-delta, as one would expect in the thermodynamic limit.
Conversely, the background component appears much closer to the asymptotic limit: 
in the inset of Fig~\ref{fig.ed} we plot the corresponding \textit{mass} distribution $P(c)$
and we superpose
the exponential distribution, which is the expected distribution for the background in the limit $N\to\infty$.
We obtain a good superposition over many orders of magnitude along the vertical direction. 

\begin{figure}[ht!]
\includegraphics[width=0.46\textwidth,clip]{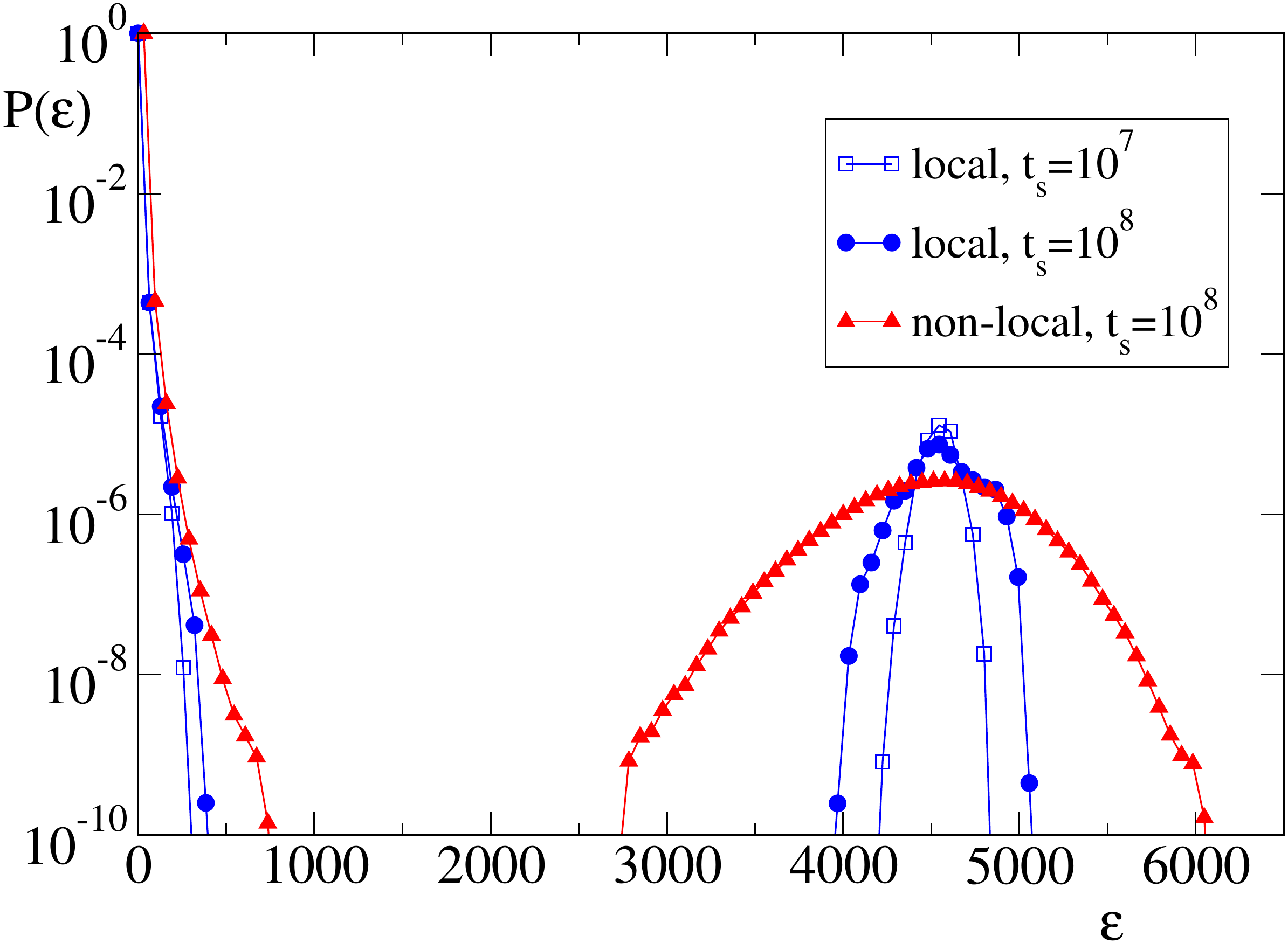}
\caption{Energy distributions for $N=25600$ and $h=2.2$ sampled over different times (see legend) for local (blue squares and circles) 
and nonlocal (red triangles) dynamics. }
\label{fig.ed_lnl}
\end{figure}

All curves of Fig.~\ref{fig.ed} are obtained by sampling the lattice configurations over a sufficiently long time $t_{s}$ to ensure stationarity.
In practice, we found that $t_{s} \sim 10^8$ is enough for such values of $N$, provided that nonlocal dynamics is implemented.
In fact, in agreement with the different relaxation exponents derived in Sec.~\ref{sec.theory}, we obtain that local
dynamics provides a substantially slower relaxation to equilibrium.
The impact of such a slowing down mechanism on energy distributions is illustrated in Fig.~\ref{fig.ed_lnl},
 where we compare the stationary energy distribution for
$N=25600$ and $h=2.2$ obtained for sampling times $t_{s}=10^8$ and nonlocal dynamics, with 
the analogous distribution corresponding to local dynamics. In the latter case, two different sampling times are considered: 
   $t_{s}=10^7$ and $t_{s}=10^8$. Noticeably, for local dynamics the whole energy distribution of the condensate is still far from
the stationary one. On the other hand, the background component is found to be more robust against $t_s$   and it appears to be
 modified only in the tails of the distribution. These observations indicate  that in the condensed phase,
  local and nonlocal dynamics differ  predominantly in the configurations of the condensate component, while they 
  behave more similarly for the background one.

\subsection{Far-from-equilibrium relaxation dynamics}
\label{ss.rd}
In the previous subsection we have put in evidence some findings 
 that are important to understand the coarsening dynamics discussed hereafter.
(i)~
For large but finite lattice sizes the equilibrium energy distribution is far from being representable as a homogeneous background plus a single peak.
This is true in the thermodynamic limit, but our results show that 
sizes of orders of tens of thousands of sites are still far from such limit.
(ii)~Local and nonlocal dynamics are different, the latter one being more effective to attain equilibrium.
(iii)~Before attaining equilibrium, local dynamics provides an energy distribution of the condensate which is
narrower than the equilibrium one.

In this subsection we show how the analytic predictions on the coarsening process discussed in Sec.~\ref{sec.theory} are modified when
the evolution for finite times and finite sizes is considered.
\begin{figure}[h!]
\includegraphics[width=0.46\textwidth,clip]{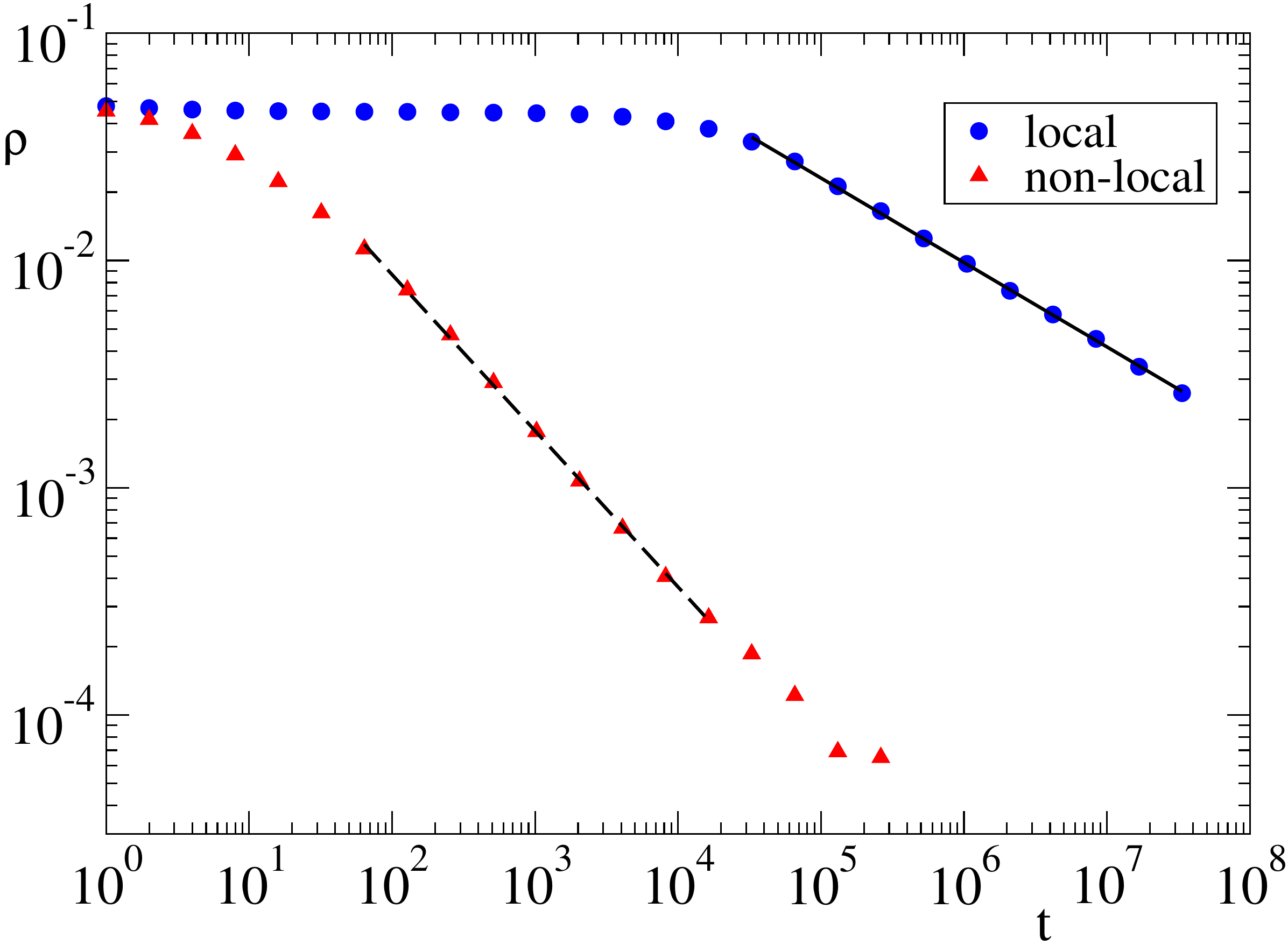}
\caption{Time dependence of the peak density for local (blue circles, $N=1600$) and nonlocal (red triangles, $N=51200$) dynamics.
Simulations refer to $h=7.6$.
Straight and dashed lines are fits to the power-law regimes, providing effective coarsening exponents
$n\simeq 0.37$ for local dynamics and $n\simeq 0.68$ for nonlocal dynamics. 
}
\label{fig.coars_lnl}
\end{figure}
In Fig.~\ref{fig.coars_lnl} we illustrate the evolution of the density of peaks $\rho(t)$ during the
coarsening process  for local and nonlocal dynamics.
In this setup, the system is initialized in a homogeneous state with a large number of small peaks.
We also choose a relatively large energy density $(h=7.6)$, which implies well separated
backgorund and peaks distributions close to the equilibrium localized  state. 
As expected, the nonlocal dynamics provides a much faster relaxation to equilibrium than the local one.
This is a consequence of the larger relaxation exponent in the regime where the power-law decay of $\rho(t)$
is established, but also of the shorter transient that precedes the coarsening regime.
The different duration of the transient stage in the two cases is 
ascribable to the details of the
kinetics: for local dynamics, peaks indirectly interact after a finite time $\Delta t\sim \ell$, while 
for nonlocal interactions peaks can directly  exchange energy  after  $\Delta t\sim 1$.
 
\begin{figure}[h!]
\includegraphics[width=0.46\textwidth,clip]{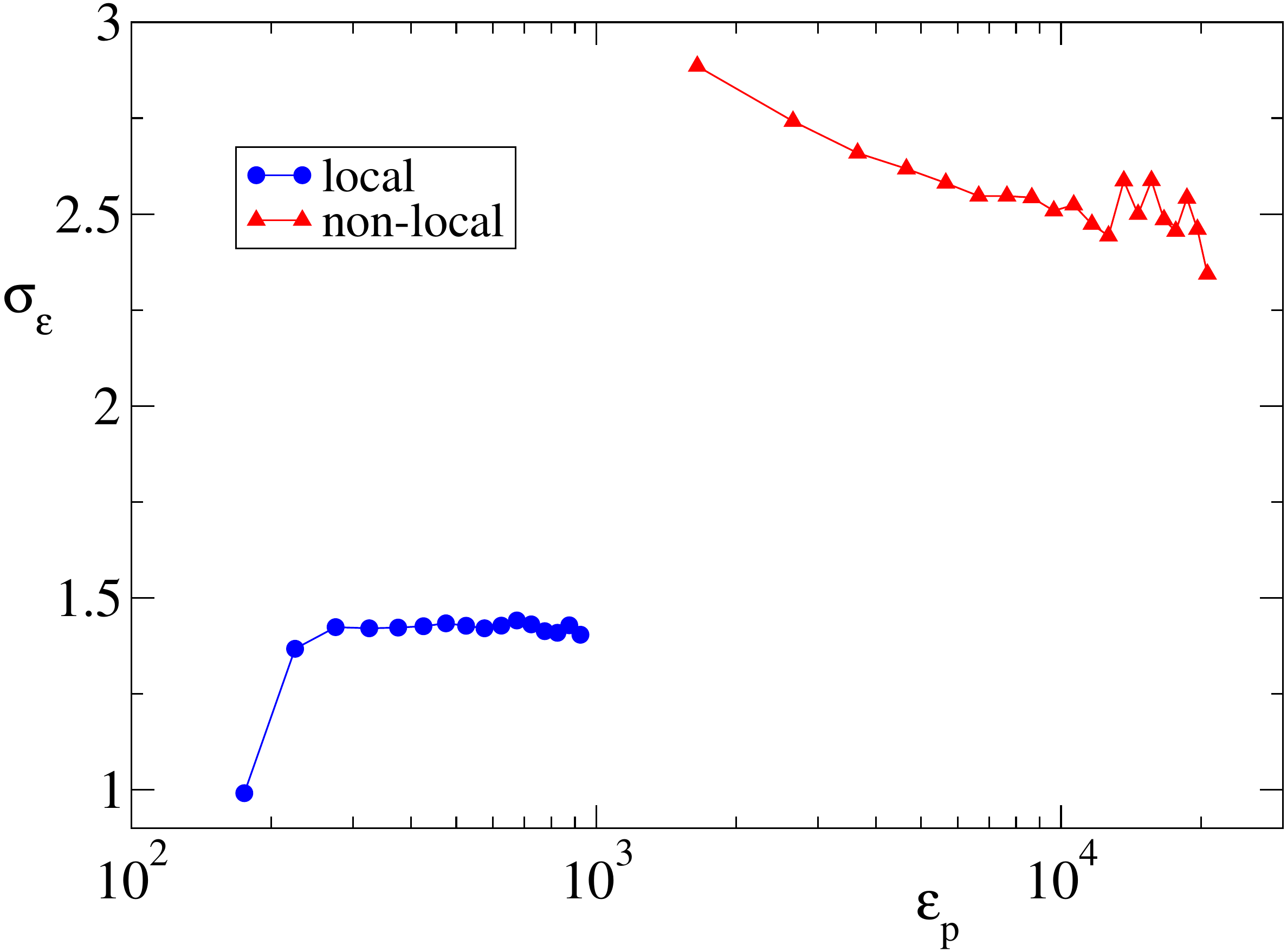}
\caption{
Peaks  energy exchanges $\sigma_\e$  as a function of the peak energy $\e_p$
for local (blue circles) and nonlocal (red triangles) dynamics. Same parameters of Fig.~\ref{fig.coars_lnl}.}
\label{fig.dE_lnl}
\end{figure}

 On the other hand, the relaxation exponents estimated from Fig.~\ref{fig.coars_lnl} are larger than the analytic ones, which 
 were obtained by first taking the limit $N\to\infty$ and then the limit $t\to\infty$.
In fact, we get $n\simeq 0.37$ instead of $n=1/3$ for local dynamics and $n\simeq 0.68$ instead that $n=1/2$ for
nonlocal dynamics.
It is worth noting that the  exponent found numerically for local dynamics is closer to the analytic one
than that for nonlocal dynamics. 
We interpret this effect again as a consequence of the different behavior of the two dynamics on a generic
multi-peak state occurring for  finite sizes and times.
Indeed, nonlocal dynamics allows a couple of peaks to directly interact, while the exponents $n=1/2$ and $n=1/3$ 
have been derived by assuming that a peak interacts with sites whose mass/energy is much lower, so that
the variation of the peak's energy $\delta \e_p$ after a generic microscopic update is constant, independent of $\e_p$.
The above argument is corroborated by Fig.~\ref{fig.dE_lnl}, where we plot  the 
standard deviation of the peak energy variations, $\sigma_\e= \langle\, (\delta \e_p)^2 \, \rangle^{1/2} $,
versus the peak energy itself. In the presence of local dynamics there is a clear regime  in which $\sigma_\e$
is constant, while this is not true for nonlocal dynamics.
We remark that the presence of peaks of different heights does not affect Fig.~\ref{fig.dE_lnl} if dynamics is local
because each peak only interacts with the background while within nonlocal dynamics peaks interact with each other
and energy variation do depend on peaks' energy.

Concerning the residual discrepancy on the coarsening exponent found in the case of
local dynamics, we have not found clear indications on its dependence
on the system size.
As one can deduce from Fig.~\ref{fig.coars_lnl}, this type of analysis is computationally challenging because of the overall slowness of the 
relaxation process and the resulting need to sample the dynamics over extremely long time scales.

\section{Conclusions}
\label{sec.conc}

The relevance of finite-size effects in statistical mechanics and condensed matter physics is not limited to the realm of microscopic systems
with few degrees of freedom.
Several examples, as those found in heat transport problems~\cite{lepri20,livi22},
 clarify that even for very large system sizes the behavior
of the system might be substantially different from what expected in the thermodynamic limit. 
In this paper we have shown that a simple model of real-space condensation displays persisting
temporal and spatial finite-size effects.
We have analyzed their impact both on equilibrium properties in the condensed phase and on the relaxation, far from equilibrium coarsening dynamics.
As customary in the absence of a fully macroscopic regime, finite-size effects are found to crucially depend on the properties of the microscopic dynamics. Indeed
we have shown that local and nonlocal update rules undergo qualitatively different finite-size corrections, which are
nontrivially interlaced with finite-time effects. In this respect, we have found that the nonlocal dynamics allows for a faster 
relaxation to equilibrium than the local one. On the other hand, the corresponding coarsening law is more affected 
by direct multi-peak interactions and the coarsening exponent exhibits evident deviations  from the single-peak picture.

The asymptotic equilibrium behavior, i.e. the thermodynamic limit $N\to\infty$, raises no doubts: the equilibrium 
configuration is a single peak on top of a uniform background, corresponding to an energy distribution $P(\epsilon)$ having 
a Dirac-delta peak at $\epsilon=(h_c -h)N$. What we have shown and discussed, see Figs.~\ref{fig.ed} and \ref{fig.ed_lnl},
is that extremely large sizes and times are necessary to observe it. We can even give a rough evaluation of them.
$N^*$ scales as $(h-h_c)^{-3}$~\cite{GIP21},
we need to consider system sizes $N\gg N^*$ to observe the asymptotic behavior, and
the relaxation of a system of size $N$ requires at least a time of order $N^3$ (local dynamics) or $N^2$ (global dynamics).
Therefore, in order to attain equilibrium in systems where the thermodynamic limit is visible we need a time which
grows at least as $1/(h-h_c)^6$ when we approach the critical point. 

The asymptotic limit in relaxation dynamics is more subtle. 
Since the coarsening exponent is evaluated from simulations where the average distance $L=1/\rho$ between
peaks varies in an interval $(L_1,L_2)$, in order to observe, e.g., the expected \textit{local} 
value $n=1/3$ two conditions are necessary: first, the energy distribution must be the expected one
for the thermodynamic limit (uniform background plus a single peak) already at system's sizes of
order $L_1$, which may be very binding if we are close to the condensation threshold (see previous
paragraph); second, $L_2$ must be at least one order of magnitude larger than $L_1$ and $N$ must be significantly
larger than $L_2$.

Even if above conditions are all satisfied, nonlocal dynamics puts additional caveats in order to
observe the \textit{nonlocal} exponent $n=1/2$, because of the possibility of a direct interaction among
peaks. Is such interaction negligible in the thermodynamic limit? If so, we expect that the upper curve (red triangles)
in Fig.~\ref{fig.dE_lnl} asymptotically attains the value $\sigma_\epsilon \simeq 1.4$, corresponding to the
energy exchange of the local model (blue circles). 
It is clear that for this to happen (if it happens) it is necessary that $\epsilon_p$
and therefore $N$ are much larger than the values that can be investigated in reasonable times.

\begin{acknowledgments}
PP acknowledges support from the MIUR PRIN 2017 project 201798CZLJ.
\end{acknowledgments}

%\bibliography{condensation}


\end{document}